# Magnetic field effects on dynamics of the ethylammonium nitrate ionic liquid confined between polar glass plates

Andrei Filippov[1,2,]*

[1]*Chemistry of Interfaces, Luleå University of Technology, SE-97187 Luleå, Sweden*
[2]*Institute of Physics, Kazan Federal University, 420008 Kazan, Russia*

*Andrei.Filippov@ltu.se

Self-diffusion and NMR relaxation of ethylammonium (EA) cations were studied in the protic ionic liquid ethylammonium nitrate (EAN) confined between parallel polar glass plates separated by a few µm. Samples were studied by $^1$H NMR at 293 K as a function of time after placement in a static magnetic field of 9.40 T (a 400 MHz NMR spectrometer). Immediately after sample placement, the diffusion coefficient of EA ($D$) was measured to be 7 $10^{-11}$ m$^2$/s, which was by a factor of ~2 larger than that in bulk EAN, according to previously reported data (http://arxiv.org/abs/1703.00233). In addition, the transverse NMR relaxation $T_2$ of –NH$_3$ protons was shorter by a factor of ~22 in comparison with that in bulk EAN. Further exposure to the static magnetic field at this temperature leads to gradual changes of $D$, $T_1$ and $T_2$ with a time constant (change by a factor of "e") of ~70 min and a total time to reach new equilibrium values for $D$, $T_1$ and $T_2$ of longer than 240 minutes (4 h). This process does not depend on the orientation of the glass plates relative to the magnetic field. Removing the sample from the magnetic field and repeating the experiment demonstrated that complete "recovery" of the sample to the "accelerated" $D$ and shortened $T_2$ occurs in approximately 1500 minutes (*ca* 25 h). Thus, EA cation dynamics is accelerated relative to bulk when the cation is confined between polar glass plates and stored in the absence of a strong magnetic field, but demonstrates a trend to change to bulk values of $D$ and $T_2$ after exposure to a significantly strong magnetic field. Because the observed characteristic times of the change far exceed characteristic times of molecular processes in EAN, we relate this phenomenon to reversible phase transformations occurring with EAN between polar glass plates outside the strong static magnetic field, and reverse change of the phases after placing the sample in the static magnetic field. This conclusion was also supported by data from Raman spectroscopy.

**Keywords:** *Nuclear magnetic resonance; Pulsed-field gradient, Self-diffusion; NMR relaxation; Ion dynamics, Static magnetic field effects, Raman spectroscopy*

## 1. Introduction

Ionic liquids (ILs) are molten salts prepared from organic cations and either organic or inorganic anions.[1,2] Their applications are continuously expanding, including uses, for instance, as electrolyte materials in lithium batteries[3] and ultracapacitors,[4] as media for chemical reactions and separation,[2,5] and as lubricants.[6] Ethylammonium nitrate (EAN), first synthesised by Paul Walden in 1914,[7] is the most frequently reported protic IL.[1] It is used as a medium for chemical reactions, as a precipitating agent for protein separation,[5] an electrically conductive solvent in electrochemistry,[3] and in other applications. Like water, EAN has a three-dimensional hydrogen-bonding network and can be used as an amphiphilic self-assembly medium.[8]

Ions of ILs can participate in a variety of interactions that may lead to formation of ordered self-assembled structures.[9] External conditions, such as temperature, pressure, electric field and confinement, may change the structural and dynamic parameters of ILs.[9,10-13] Confinement in mesoporous carbon[10] and silicon[11,12] reportedly increases the diffusion of ILs by a factor of 2-3 or more. The latter has been explained as a consequence of decreasing the packing density of ions in pores with diameters comparable to the sizes of the ions.[11] Recently, enhanced diffusion of EA cations has been observed for EAN confined between polar glass plates separated by 4 μm,[14] which has been related to the transformation of a sponge-like structure proper to bulk EAN in a more isotropic phase. Magnetic field strength is also an important external variable that influences the properties of some organic substances, such as liquid crystals[15] and organic-based semiconductors.[16] No perceptible magnetic field effects have been observed for ionic liquids until now.[9,17]

In this communication, we report the effect of applying a strong magnetic field (in an NMR spectrometer) to EAN confined between polar glass plates separated by ~4 μm. In comparison to bulk EAN, this ionic liquid in confinement demonstrated accelerated translated diffusion of EA cations and accelerated transverse NMR relaxation rates of EA –$NH_3$ protons.[14] After subjecting this system to a strong magnetic field (9.4 T), we observed a slow decrease in the diffusion coefficient of the EA cation and a slow increase of $T_2$ NMR relaxation times for EA –$NH_3$ protons, both values approaching those in bulk EAN.

## 2. Experimental Part

EAN, prepared from an equimolar mixture of ethylamine and nitric acid, after a proton transfer consists of an ethylammonium (EA) cation and a nitrate anion (see Fig. 1). EAN used in our experiment was synthesised, pre-treated and characterised as described previously.[14]

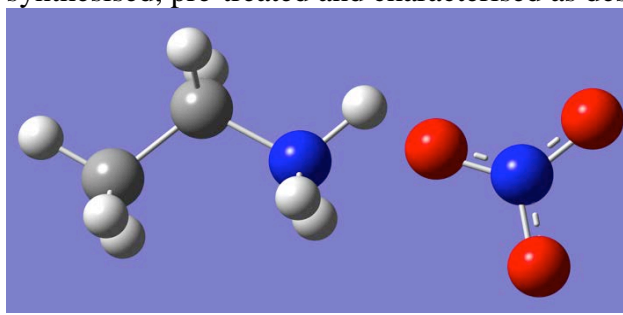

**Figure 1.** Colour model of the chemical structure of EAN, consisting of an ethylammonium cation and a nitrate anion, with nitrogen (blue), oxygen (red), carbon (dark grey) and hydrogen (light grey) atoms.

Bulk EAN is a low-viscosity liquid at room temperature. NMR measurements for bulk EAN were performed by placing it in a standard 5-mm NMR glass tube. Confined EAN was prepared with thin glass plates arranged in a stack (see Fig. 2).

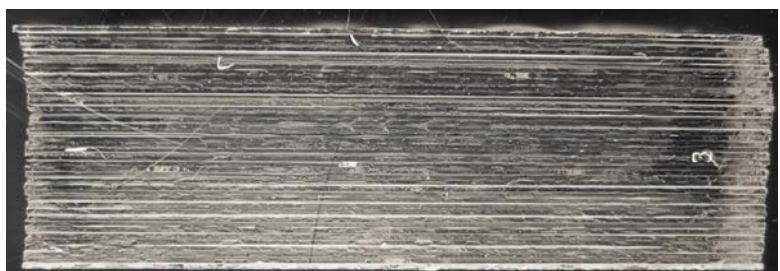

**Figure 2.** Alignment of glass plates with confined EAN.

The plates (5 x 14 x 0.1 mm, Thermo Scientific MenzelGläser, Menzel GmbH, Germany) were carefully cleaned before use. Contact angle measurements with Milli-Q water provided a contact angle near 0°, indicating hydrophilic glass surfaces. A stack of glass plates filled with EAN was prepared in a glove box in a dry nitrogen atmosphere. A drop of EAN was placed on the first plate, which was covered by the second plate onto which another drop of EAN was placed, and so on until the thickness of the stack reached approximately 5 mm. Excess EAN from the sides of the stack was removed by wiping. The plates were thereafter placed in a sample cell of square cross section. The mean distance (spacing) between glass plates was estimated by weighing the introduced EAN, which yielded $d$ ~3.8-4.5 µm. Measuring the thickness directly and subtracting from it the total glass thickness indicates that this is a consistent value. A detailed description of sample preparation and characterisation has been reported previously.[14] The spacing between the glass plates was estimated as ~4.1 µm from a series of $^1$H NMR diffusion measurements on EAN in the direction normal to the plates, performed in the diffusion time range of 3-300 ms.[14]

$^1$H NMR self-diffusion, $T_1$ and $T_2$ relaxation measurements of EAN, in bulk and confined between glass plates, were performed with a Bruker Ascend Aeon WB 400 (BrukerBioSpin AG, Fällanden, Switzerland) NMR spectrometer with a working frequency of 400.27 MHz for $^1$H, magnetic field strength of 9.4 T with magnetic field homogeneity better than 4.9·10$^{-7}$ T, using a Diff50 Pulsed-Field-Gradient (PFG) probe. Data were processed using Bruker Topspin 3.5 software. A NMR solenoid $^1$H insert was used to macroscopically align the plates of the sample stack at 0 and 90 degrees with respect to the magnetic field. The diffusional decays (DD) were recorded using the stimulated echo pulse sequence. For single-component isotropic diffusion, the signal intensity $A$ changes as:[18,19]

$$A(\tau,\tau_1,g,\delta) \propto \exp\left(-\frac{2\tau}{T_2}-\frac{\tau_1}{T_1}\right)\exp\left(-\gamma^2\delta^2 g^2 D t_d\right) \quad (1)$$

where $T_1$ and $T_2$ are the spin-lattice and the spin-spin relaxation times, respectively; $\tau$ and $\tau_1$ are time intervals in the pulse sequence; $\gamma$ is the gyromagnetic ratio for protons; $g$ and $\delta$ are the amplitude and the duration of the gradient pulse; $t_d = (\Delta - \delta/3)$ is the diffusion time; $\Delta = (\tau + \tau_1)$; and $D$ is the self-diffusion coefficient. In the experiments the gradient amplitude, $g$, was varied, while other parameters were kept constant. There is no effect of restrictions on $D$ in the direction along the plates, while for diffusion normal to the plates the effect of the plates on $D$ is noticeable at $t_d$ longer than 3 ms and increases as $t_d$ increases.[14] For diffusion in bulk and along the plates, $D$ values were obtained by fitting Eq. 1 to the experimental decays. For decays obtained in the direction normal to the plates at $t_d = 3$ ms, $D$ was calculated from the equation:

$$D_{av} = \frac{-\partial A(\gamma^2\delta^2 g^2 t_d)}{\partial(\gamma^2\delta^2 g^2 t_d)}\bigg|_{(\gamma^2\delta^2 g^2 t_d)\to 0} \quad (2)$$

All measurements were performed at 293 K and were started immediately after placing the sample in the NMR probe. The time required for a single NMR diffusion measurement was around 30 s, while that of a relaxation measurement was around 3 min.

$^1$H $T_1$ and $T_2$ NMR relaxation time measurements were performed with inversion-recovery (180°-τ-90°-fid) and CPMG (90°-τ-180°-τ-echo) pulse sequences, respectively.[20]

### 3. Results and Discussion

Placement of bulk EAN in the magnetic field of the NMR spectrometer and systematic measurements of $D$, $T_1$ and $T_2$ over 24 hours showed no changes to these parameters. For EAN confined between glass plates, we observed slow changes (on the order of tens of minutes) in the $D$, $T_1$ and $T_2$ of different chemical groups of EA, as shown in Fig. 3. Just after placing the sample in the NMR probe, $D$ (denoted as $D^*$) was measured to be a factor of ~2 higher than that in bulk ($D_0$). $T_1$ relaxation decays for protons of different groups demonstrated forms close to the exponential ones. $T_2$ relaxation for the $-NH_3^+$ group was exponential, while $T_2$ relaxation of $-CH_2-$ and $-CH_3$ groups were close to the sum of the two exponential functions. In this case, the mean values of $T_2$ were calculated as: $T_2 = (1/T_{21} + 1/T_{22})^{-1}$. Initial values of $T_2$ (just after placement of the sample in the magnetic field) of different groups are decreased by factors of ~8.8 ($-CH_2-$), ~12.1 ($-CH_3$) and ~22 ($-NH_3^+$) relative to their bulk values. Further, we observed that $D$ gradually decreased from $D^*$ to smaller values gradually approaching $D_0$ (Fig. 3A). Simultaneously, the $T_1$ of different protons decreased by a factor of ~1.25 (Fig. 3B). Mean values of $T_2$ for $-CH_2-$ and $-CH_3$ groups demonstrated small changes, while $T_2$ ($NH_3$) increased by a factor of ~10 after ~240 minutes. A similar effect was observed for the sample oriented along and normal to the glass plates relative to the magnetic field.

From Fig. 3, we see that the decrease in $D$ and the increase in $T_2(NH_3)$ reach plateaus after 4 hours. The process can be characterised by the time during which the initial values of $D$ and $T_2(NH_3)$ change by a factor of "$e$", which was estimated from Fig. 3 to be nearly 70 min for both these parameters.

The experiment was repeated several times with different time intervals, $TE$, between completion of the change in $D$ and $T_2(NH_3)$ values in the previous experiment and the beginning of the next experiment. During $TE$ the sample was removed from the magnet and stored at 293 K in the magnetic field of Earth (*ca* 0.5 $10^{-4}$ T). In cases when $TE$ was just few hours, the measured $D$ value in the beginning of the new experiment was less than $D^*$ but larger than $D_0$ and $T_2(NH_3)$ was higher than $T_2^*(NH_3)$. But in cases when $TE$ was on the order of a day or longer, $D$ and $T_2(NH_3)$ were completely "recovered" to $D^*$ and to $T_2^*(NH_3)$, respectively.

Two additional experiments were performed to ensure that the effects observed for $D$ and $T_2(NH_3)$ are due to the presence of a strong magnetic field, but not associated with other influences that may occur in the NMR spectrometer during diffusion and relaxation experiments. In the first experiment, a fully "recovered" sample was placed in the spectrometer probe for a rather long time, but no radio frequency pulses were applied during this time interval. The first measurements of $D$ and $T_2$ were performed after 24 hours and yielded the same values as those from experiments where continuous measurements were taken (i.e. with values of parameters as in Fig. 3 after 240 minutes). In another experiment, a sample that had "recovered" outside of the magnetic field of the spectrometer was placed between the poles of Neodymium permanent disk-shaped magnets with a magnetic field strength of ~0.68 T for 24 hours before placing the sample in the NMR spectrometer for measurements. After placing the sample in the spectrometer, the measured value of $D$ was between $D^*$ and $D_0$. Therefore, the presence of a strong magnetic field is the main factor influencing the $D$ and $T_2(NH_3)$ processes that occur with EAN confined between polar plates.

Neither effect, acceleration of EA dynamics nor its change under the influence of a magnetic field, are observed for bulk EAN. The time-scale of these changes is on the order of a few hours, which is much longer than the time-scale of dynamic processes occurring with individual EAN ions confined between glass plates. Indeed, the longer of the dynamic processes is the self-diffusion of the ions, which covers the inter-planar distance of nearly 100 ms. For this reason, we relate the processes occurring with EAN between glass plates to reversible phase transformations of EAN. It has been suggested that EAN, because of the competition of the strong hydrogen bonding and electrostatic interactions of the EA cation and nitrate anion, and the hydrophobic interactions of $-CH_2-$ and $-CH_3$ groups of the EA cation, may form a bi-continuous sponge-like structure.[9,21-23]

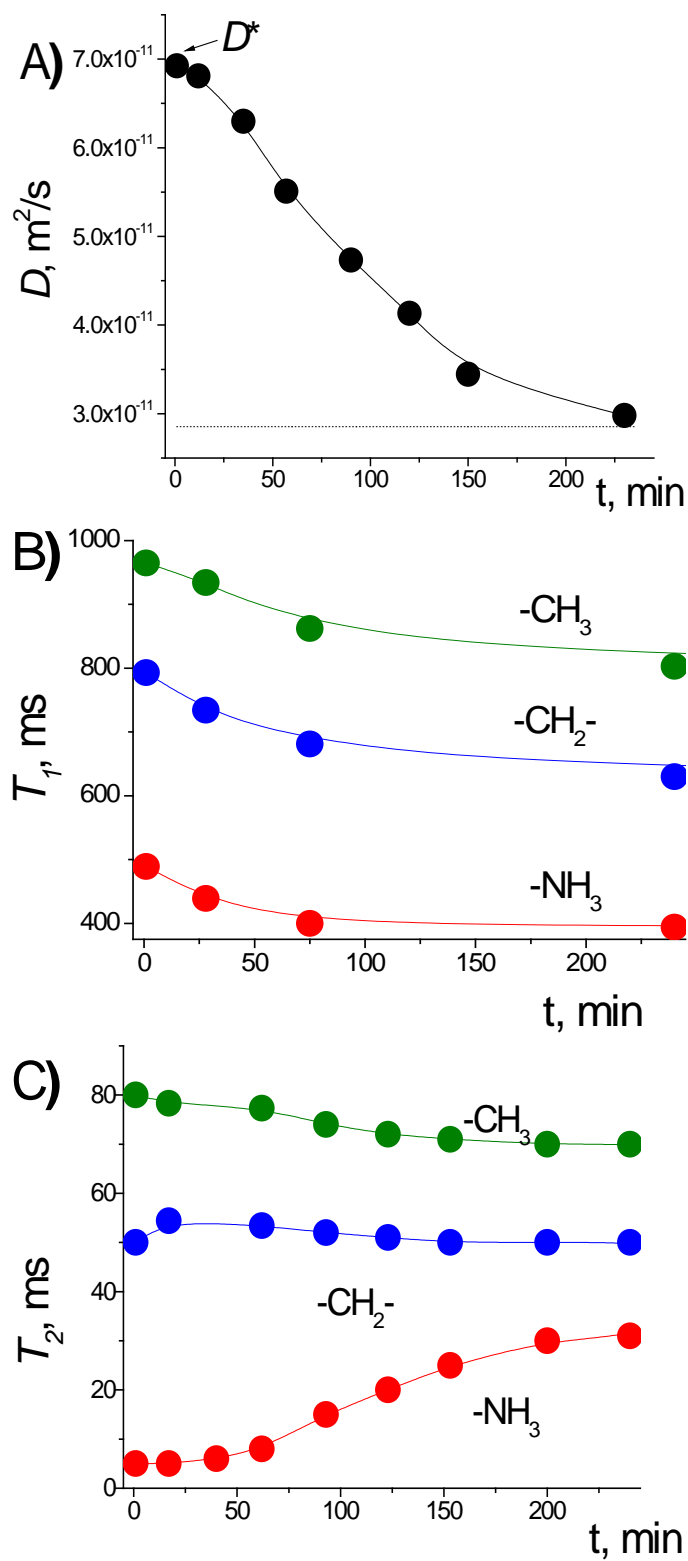

**Figure 3.** Change of **A**) Diffusion coefficient of EA cation, **B**) $T_1$ NMR relaxation times and **C**) $T_2$ NMR relaxation times of protons of different chemical groups of the EA cation after placing the EAN confined between polar plates in a magnetic field of 9.4 T. $D$ was measured along the plates, $t_d = 3$ ms.

Because this structure is characteristically a balance of forces, structural parameters can vary as a consequence of changes in temperature, pressure and chemical modification of ions.

Hindered diffusion is typical for this structure, as well as for the bi-continuous cubic phase, because of the curvilinear trajectories of ions along the micro-phase borders.[24] It could be suggested that the balance of forces and structural parameters of the sponge-like phase may be modified in the presence of additional interactions of ions with the polar surfaces of glass plates.

The change in structural parameters enhances the mobility of the ions.[25] Another possibility is the transformation of the sponge-like phase to a more isotropic phase. The phase transformation leads to enhancement for diffusive transport and for accelerated -$NH_3$ proton exchange. Application of a strong magnetic field additionally contributes to inter-ion interactions, which might be more efficient for moving ions, leading to further disturbance of the existing equilibrium. Thus, the application of a strong magnetic field to EAN confined between polar plates allows one to control the dynamics of EA ions.

To check the hypothesis of formation of a new phase of EAN in the confinements and the destructive effects of a strong static magnetic field on this phase, we performed Raman spectroscopic measurements for bulk EAN and EAN confined between the polar glass plates. The Raman spectra were collected with a Ramanscope III spectrometer combined with a SENTERRA module. The spectrometer was equipped with a 532 nm excitation laser for use in the standard normal incidence sampling geometry. The data were accumulated in the range from 30 to 3710 $cm^{-1}$ using an incident laser power of about 20 mW. An integration time of 5 s and 90 sample scans were used to acquire the spectra. Background spectra from glass plates with the same geometry but without EAN were subtracted. Fig. 4 summarises Raman spectra for bulk EAN and EAN in confinement with and without exposure of the sample to the static magnetic field (MF) of 9.4 T.

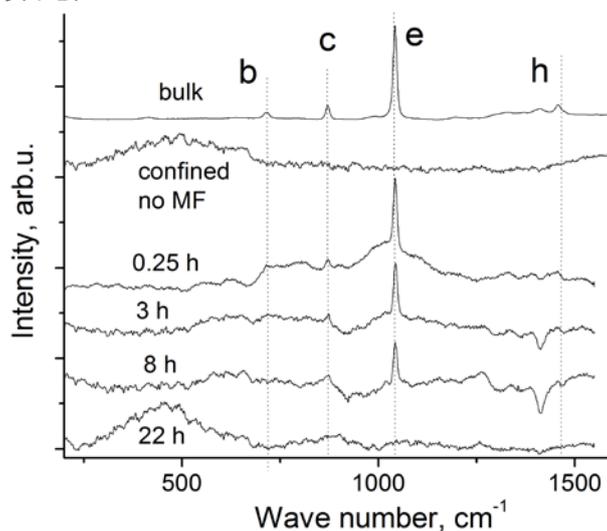

**Figure 4.** Low-frequency modes of Raman spectra of bulk EAN, EAN confined between glass plates before placing in magnetic field (MF) of 9.4 T and EAN confined between glass plates after consequent exposure in the magnetic field of 9.4 T during 12 hours with the following "recovery" in the magnetic field of Earth, as a function of the "recovery" time (0.25, 3, 8 and 22 hours).

The assignment of lines for the Raman spectrum of bulk EAN was taken from literature.[26,27] Most intensive lines are due to: **b** – bending $NO_3^-$, **c** – $C_n$-N symmetric stretching, **e** – $NO_3^-$ symmetric stretching motion and **h** – -$CH_2$ and -$CH_3$ bending motion. Confinement of EAN significantly changes its original (EAN in bulk) Raman spectrum. No narrow lines characteristic for bulk EAN were observed. The high-frequency modes detected for bulk EAN in the range 2600-3400 $cm^{-1}$ are absent in the Raman spectrum of EAN between glass plates.

Raman data are an indirect confirmation of the suggested hypothesis on a new phase of EAN in confinement, as discussed above in the analyses of diffusion and NMR relaxation data.

After exposure of the EAN sample to the external magnetic field of 9.4 T, high-frequency modes of motion do not change remaining as that before the exposure. In the low-frequency range, lines corresponding to **c** and **e** modes appear. These modes have linewidths and intensities comparable to those in bulk EAN. During "recovery" of the sample, intensities of **c** and **e** modes monotonously decrease and finally disappear after *ca* 22 hours of the "recovery" time (see Fig. 4) in a good agreement with NMR data.

In summary, we report on the first experimental observation of magnetic field effects on dynamics of a non-magnetic ionic liquid, ethylammonium nitrate. Further experimental and theoretical studies are required for understanding of a new phase of EAN stabilised between polar glass plates at room temperatures: the effect of roughness, surface polarity and changing the ionic liquid constituents. Given that the liquid properties are significantly altered, these results have strong implications for interface-intensive applications of ILs, such as their use as lubricants, in electrochemical systems, and in supported catalysis.

**Acknowledgments**

We thank Mylene Trublet and Allan Holmgren for acquisition of Raman spectra and useful discussions. The Foundation in memory of J. C. and Seth M. Kempe and the laboratory fund at LTU are gratefully acknowledged for providing grants, with which the Bruker Aeon/Avance III NMR spectrometer at LTU was purchased. "Scriptia Academic Editing" is acknowledged for proof-reading of this manuscript.